\documentclass[lettersize,journal]{IEEEtran}
\usepackage{amsmath,amsfonts}
\usepackage{algorithmic}
\usepackage{algorithm}
\usepackage{array}
\usepackage[caption=false,font=normalsize,labelfont=sf,textfont=sf]{subfig}
\usepackage{textcomp}
\usepackage{stfloats}
\usepackage{url}
\usepackage{verbatim}
\usepackage{graphicx}
\usepackage{multirow}
\usepackage[nolist]{acronym}
\usepackage{cite}
\usepackage{tabularray}
\usepackage{orcidlink}
\usepackage{svg}
\begin{acronym}
    \acro{DoF}{Depth-of-Field}
    \acro{FP}{Full Parallax}
    \acro{DSLF}{Densely Sampled Light Field}
    \acro{DASC}{DoF-Aware Scene Complexity}
     \acro{SI}{Spatial Information} 
     \acro{CF}{Colorfulness}
     \acro{TI}{Temporal Information}
     \acro{PMF}{Probability Mass Function} 
     \acro{HMD}{Head Mounted Display}
     \acro{LRT}{Likelihood Ratio Test}
\end{acronym}

\DeclareMathOperator{\EX}{\mathbb{E}}% expected value

\begin{document}
\bstctlcite{IEEEexample:BSTcontrol}
\title{
%Perceptual Assessment of Depth-of-Field Visualization in Light Field Displays

%erceptual Depth-of-Field Aware Scene Characterization for 3D Visualization on Light Field Display

%Perceptually aware scene characterization for 3D visualization on light field display

%Depth-of-Field-Aware Scene Characterization for 3D Light Field Visualization

DASC: Depth-of-Field Aware Scene Complexity Metric for 3D Visualization on Light Field Display

}

\author{Kamran Akbar\,\orcidlink{0009-0001-8705-8808} \thanks{Kamran Akbar and Robert Bregovic are with the with the Faculty of Information Technology and Communication Sciences, Tampere University, Kalevantie 4, 33100 Tampere, Finland. (e-mail: \href{mailto:kamran.akbar@tuni.fi}{kamran.akbar@tuni.fi}, \href{mailto:robert.bregovic@tuni.fi}{robert.bregovic@tuni.fi})}, Federica Battisti\,\orcidlink{0000-0002-0846-5879} \thanks{Federica Battisti is with the Department of Information Engineering, University of Padova, Via Gradenigo 6/b, 35131, Padova, Italy. (e-mail: \href{mailto:federica.battisti@unipd.it}{federica.battisti@unipd.it}).}, Robert Bregovic\,\orcidlink{0000-0002-3878-7588}}

\markboth{Journal of IEEE Transaction on Multimedia}%
{Shell \MakeLowercase{\textit{et al.}}: A Sample Article Using IEEEtran.cls for IEEE Journals}


\maketitle

\begin{abstract}
Light field display is one of the technologies providing 3D immersive visualization. However, a light field display generates only a limited number of light rays, which results in finite angular and spatial resolutions. Therefore, 3D content can be shown with high quality only within a narrow depth range, referred to as \ac{DoF}, around the display screen. Outside this range, due to the appearance of aliasing artifacts, the image quality degrades proportionally to the distance from the screen. One solution to mitigate the artifacts is DoF rendering which blurs the content in the distorted regions, but can result in the removal of scene details. With the aim of determining the user preferred blurring level, in this research we proposed a \ac{DASC} metric that characterizes a 3D content based on geometrical and positional factors and the light field display's \ac{DoF}. Furthermore, we evaluated the observers' preference across different levels of blurriness caused by \ac{DoF} rendering ranging from sharp, aliased scenes to overly smoothed alias-free scenes. We conducted this study over multiple custom-designed scenes that provide representation of different content types. Based on the outcome of subjective studies, we proposed a model that takes the value of \ac{DASC} metric as input and predicts the preferred blurring level for a given scene as output.

\end{abstract}

\begin{IEEEkeywords}
Light field display, angular resolution, spatial resolution, depth-of-field rendering, aliasing artifact, subjective study.
\end{IEEEkeywords}

\section{Introduction}
\label{introduction}
In recent years, 3D visual media made a significant progress in delivering immersive experiences to a broader audience. This trend has been pushed by a number of applications that span from entertainment to training and education. A key aspect that supports this success is the technological development of 3D visualization that allows the users to perceive an increased feeling of immersion in the content that is displayed. Various methods exist for 3D visualization, and among these, light field displays exploit the concept of light field to visualize the underlying 3D scene \cite{levoy_LightField}.

The current generation of light field displays reproduces only a finite number of light rays. Therefore, the resolution of each view (spatial resolution) and the number of views visualized in the display's field of view (angular resolution) are limited. This limitation causes light field displays to have a finite depth range around the display screen in which the content can be visualized with the highest possible resolution. This range is referred to as the display's \ac{DoF}. Outside the \ac{DoF}, the resolution reproducible by the display drops with the distance from the display's screen\cite{bregovic_signal_2019} \cite{AKbar_antialiasing_LF_display} \cite{Balogh2008THEHS}. 

%As a result of limited spatial and angular resolutions, there is a possibility that the content outside the display's \ac{DoF} is visualized with distortions. 

Due to the drop in resolution, content visualized outside of the display’s DoF may exhibit distortions. The distortions are seen as ghostly images around the main objects, and are referred to as aliasing artifacts. They can be mitigated with \ac{DoF} rendering algorithms which keep the content sharp in the display's \ac{DoF} and blur everything else proportional to the distance from the display's screen plane \cite{AKbar_antialiasing_LF_display} \cite{Zwicker} \cite{Isaksen}. The quantitative relation between the amount of blur and the display's spatial and angular resolutions is discussed in \cite{Kamran4738}. Although blurring the aliased content enhances motion parallax smoothness, it results in detail loss in regions outside the display's \ac{DoF}. Therefore, subjective quality assessment can help in understanding the perceptual impact of different artifacts ranging from sharp aliased content with rough motion parallax to blurred anti-aliased content with smooth motion parallax. Based on the result of subjective studies, it is possible to define a general model that estimates the preferred blurring level of any given scene.

 Light field quality assessment is a well investigated topic in the state-of-the-art and several approaches have been presented to estimate the quality perceived by the users. Those methods evaluate features such as depth, refocusing, spatial information, contrast, and colorfulness. In most cases, these % . Gill et al. \cite{Visual_Perception} measured the visual attention and saliency of light field content using an eye tracker and a 2D display. Paudyal et al. \cite{characterization} characterized the light field content based on various features e.g. depth, refocusing, spatial, contrast, and colorfulness. Paudyal et al. \cite{Parad_Paud} studied the perceptual impact of different rendering techniques and artifacts on the QoE of light field images. Tamboli et al. \cite{TAMBOLI201642} proposed a 3D full-reference quality metric considering the spatial-angular nature of the 3D content. Adhikarla et al. \cite{Adhikarla879} and Min et al. \cite{Min2319} performed a subjective study on different artifacts resulting from compression, reconstruction, displaying, etc. happening in light field content. Tian et al. \cite{Tian387} introduced a full-reference metric for light field images based on geometry, spatial features, and colorfulness. The mentioned 
 methods estimate the quality of light field content displayed on 2D displays. However, 2D displays provide a limited feeling of immersion due to lack of depth cues and motion parallax. Therefore, this paper aims at assessing the perceptual impact of \ac{DoF} guided visualization on light field displays by considering its pros in terms of better immersive experience and its cons due to the presence of aliasing artifacts. In our research, the main contributions are: 
\begin{itemize}
    \item Proposing an objective metric that estimates geometric and position factors of the scene (Section \ref{model}).
  \item Designing and creating a dataset comprising nine \ac{FP} \acp{DSLF} from diverse scenes with objects located inside and outside the display's \ac{DoF} (Section \ref{Subjective Study}). 
  \item Designing and implementing a protocol for subjective quality assessment on a light field display to determine the perceptual preference of different blurring levels (Section \ref{Subjective Study}).
  \item Establishing a correlation model between the proposed objective metric and the opinion scores collected during the subjective study (Section \ref{evaluation}).
 % \item Proposing an approach for confirming the effectiveness of the correlation model.
\end{itemize}

The rest of the paper is organized as follows. Section 2 focuses on background information about light field, light field displays, and \ac{DoF} in light field displays. Section 3 reviews the existing research on light field quality assessment. Section 4 discusses details about the proposed scene quantification metric. Section 5 describes the process of subjective quality assessment and dataset design and creation. Section 6 presents the analysis of the experimental and validation studies, and correlates them with the values of the proposed model. Section 7 provides the concluding remarks, including research limitations, and possible future works. 

\section{Background}
\label{Background}

\subsection{Light Field and Light Field Displays}
Light field is a representation of a 3D scene that indicates the intensity of light rays coming from different positions and directions \cite{levoy_LightField}. A common way to represent a light field is by using the two plane parametrization. In this representation, each ray in the light field, $L(s, t, u, v)$ carries the light's information between two planes, i.e. camera and the image planes, where $(s, t)$ corresponds to a position on the camera plane and $(u, v)$ corresponds to a pixel coordinate on the image plane. 

Light field displays are 3D displays that visualize 3D content using the light field concept. Light field displays reproduce a finite amount of light rays, and are quantified by angular and spatial resolutions. The spatial resolution corresponds to the highest resolution of each view and the angular resolution corresponds to the number of views in the display's field of view \cite{bregovic_signal_2019} \cite{levoy_LightField}. There are both experimental and analytical methods for measuring display's angular and spatial resolutions. In analytical methods, the display's geometrical configuration is needed \cite{Zwicker} \cite{Wetzstein}, whereas experimental methods determine these resolutions by analyzing the appearance of different signals with varying frequencies on the display \cite{Kovacs} \cite{singleImage}.

\subsection{Depth-of-Field in Light Field Displays}
\begin{figure}[h!]
    \centering
    \includegraphics[width=0.8\linewidth]{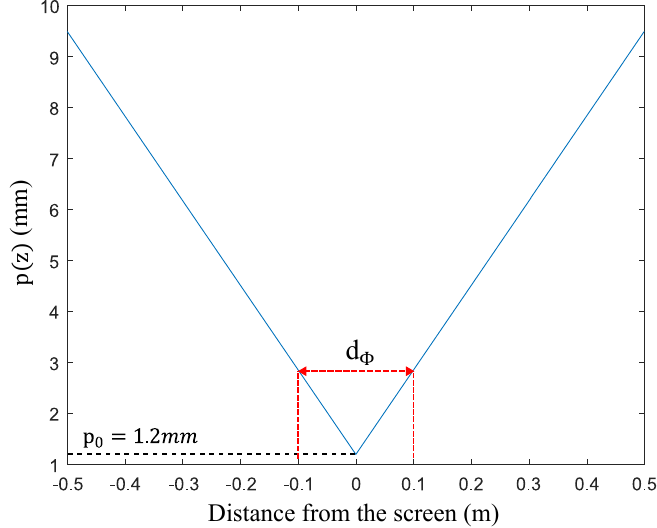}
    \caption{ The smallest reproducible feature size at a specific distance $z$ from the light field display's screen with $z_f = 3m$, $p_0 = 1.2 mm$, and $\alpha_s = 0.95^{\circ}$.}
    \label{fig:feature_size}
\end{figure}

\begin{figure}[h!]
    \centering
    \includegraphics[width=\linewidth]{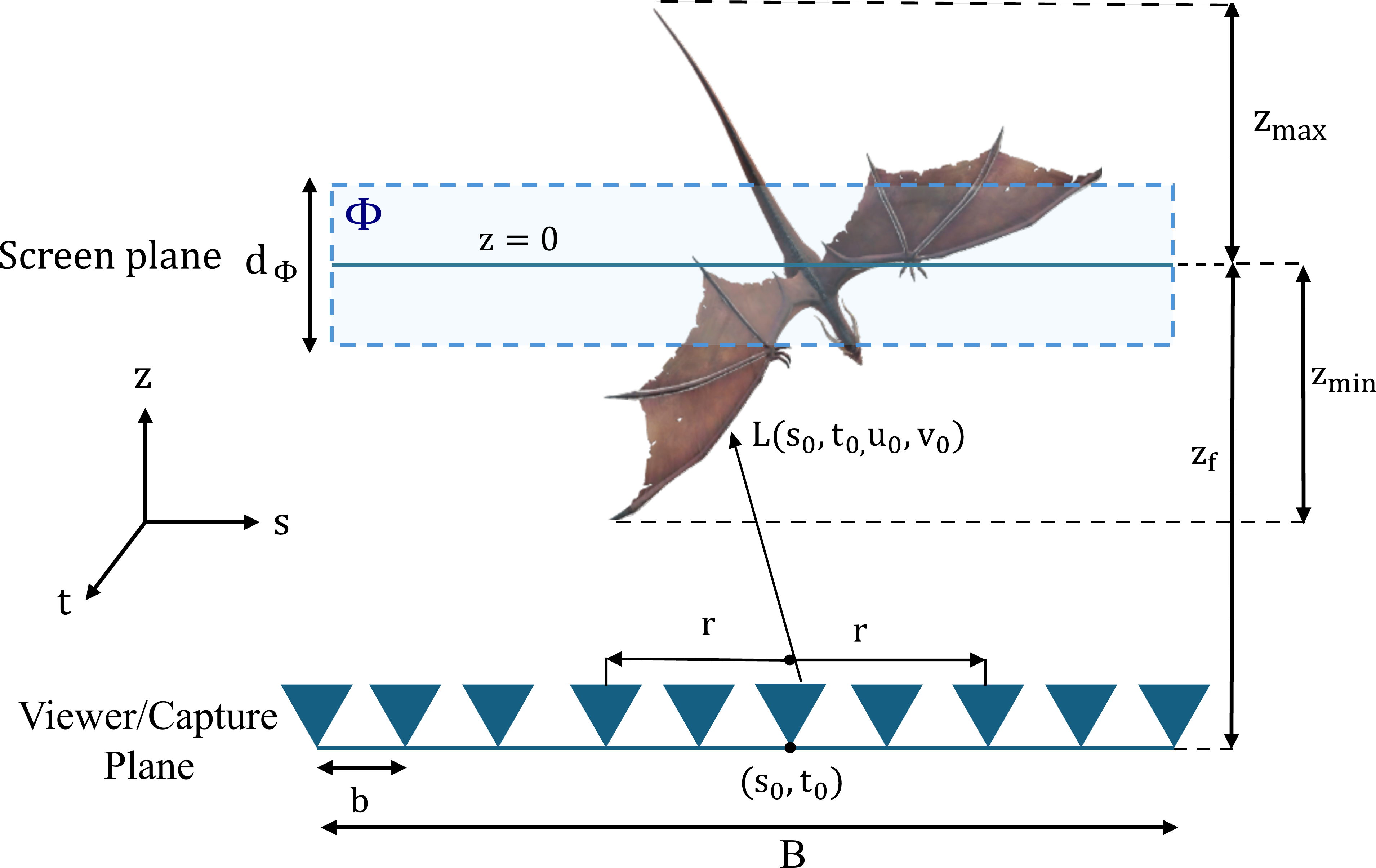}
    \caption{ Illustration of the light field rendering and visualization setup. }
    \label{fig:setup}
\end{figure}

The spatial and angular resolutions of light field displays define the limited depth range around the display screen in which the content can be visualized with the highest spatial resolution. This depth range is referred as the display's \ac{DoF}. The size of the smallest feature replicated in the display's \ac{DoF} is equal to the pixel size, $p_0$, on the display's screen \cite{Kamran4738}. However, the display's ability to reproduce the smallest feature size $p(z)$ varies with distance from the screen as shown in Figure \ref{fig:feature_size} and can be expressed according to\cite{Balogh2008THEHS} as:

\begin{gather}
\label{eq:feature_size}
p\text{(}z\text{)} = p_0+\left|z\right| \tan(\alpha_s),
\end{gather}
where $\alpha_s$ and $z$ refer to the display's angular sampling rate and the distance from the display screen, respectively. The object's distance from the front and back of the display screen is represented by the negative and positive values of $z$, respectively. The angular sampling rate $\alpha_s = \frac{FoV_s}{N_s}$ depends on the number of views $N_s$ generated in the display's field of view $FoV_s$ \footnote{Field of view in light field displays is the angular range in front of a display, where the 3D content is visualized with the correct perspective.}. In light field displays, the \ac{DoF} region $\Phi$ (as illustrated in Figure \ref{fig:setup}) corresponds to the distance range $d_\Phi$ where $p\text{(}z\text{)} \leq 2p_0$. 

%However, the intensity distribution of light rays on the light field display's screen overlaps within a narrow neighboring region, blurring features of approximately size $p_0$ \cite{Balogh2008THEHS}. Therefore, in this study we consider a larger distance range for display's \ac{DoF} $D_\Phi = 2\hat{D}_\Phi$. 

\subsection{Depth-of-Field Rendering for Light Field Displays}
Providing that part of the 3D content might be visualized outside the display's \ac{DoF}, aliasing artifacts will appear which are seen as ghostly images around the main objects in the scene. For mitigating the aliasing artifacts, in this paper, we adopt the DoF rendering method proposed in \cite{AKbar_antialiasing_LF_display}. This method requires \ac{FP} \ac{DSLF} \cite{bregovic_signal_2019}, and applies circular filters to each view of the light field, blurring the content depending on its distance from the display's screen \footnote{Each view in \ac{FP} \ac{DSLF} is recentered, as discussed in \cite{AKbar_antialiasing_LF_display}.}. The circular filter impulse response is defined as:
\begin{gather}
\label{eq:impulse_response}
h_{s_0, t_0}(s, t) = \left\{ \begin{array}{cl}
1 & : \ \text{for} \sqrt{(s-s_0)^2+(t-t_0)^2}\leq r \\
0 & : \ \text{otherwise},
\end{array} \right.
\end{gather}
where $r$ is the radius of the circular filter that determines the number of views. The optimal filtering radius $\hat{r}$ can be evaluated from the angular sampling rate of the light field display $\alpha_s$ and the \ac{DSLF} $\alpha_c$ as:

\begin{gather}
\label{eq:decimation}
\hat{r} = 0.5\left\lfloor \frac{\alpha_s}{\alpha_c} \right\rfloor \\
\alpha_c = \tan^{-1}(b/z_f),
\end{gather}
where $z_f$ is the distance of the viewer plane to the display screen, and $b$ is the distance between two adjacent views in the \ac{DSLF} \cite{bregovic_signal_2019}. The circular filter is applied to each view of the \ac{DSLF} $(s_0, t_0)$ separately, 
\begin{gather}
\label{eq:filter}
L_f(s_0,t_0,u,v) = \frac{1}{M}\sum_{s}^{}\sum_{t}^{}h_{s_0, t_0}(s, t)L(s,t,u,v),
\end{gather}
where the normalization factor $M = \sum_{s}^{}\sum_{t}^{}h_{s_0, t_0}(s, t)$ is the number of contributing views in the filtering process. The circular filtering keeps the content in the display's \ac{DoF} sharp but gradually blurs everything outside the \ac{DoF} region thus causing detail loss in the visualized content. In this research, we use circular filters with different radii to apply various levels of blurring to the light field content.

\section{Related Works}
\label{Related Works}
In this research, by means of subjective study, we evaluate the perceptual preference of 3D visualization on light field display for different blurring levels. Based on the participants' collected opinion score, we developed a metric to estimate the proper amount of blurring for any given scene.

Developing a metric for assessing light field quality is an important domain of research driven by the growing adoption of light field technology in immersive applications. Kara et al. \cite{electronics_Kara_0953} categorized research questions related to light field quality assessment into system-related, visualization-related, and viewer-related. System-related questions concern key performance indicators and technological comparisons of the displays. Visualization-related questions are associated with distortions resulting from different image processing algorithms, which influence the visualization. Viewer-related questions correspond to the assessment of the immersion, interaction, and usability of light field displays. According to Kara et al., many viewer-related research questions remain unanswered.

Gill et al. \cite{Visual_Perception} introduced visual saliencies for light field imaging by gathering eye tracking data. The main purpose was to differentiate the visual attention in light field and conventional 2D images. Battisti et al. \cite{Vis_Technique} compare various light field visualization techniques on 2D displays. Paudayal et al. \cite{characterization} proposed a method that characterizes light field data based on specific features e.g. depth, refocusing, spatial, temporal, contrast, and colorfulness. This method facilitates the light field data selection for effective perceptual assessment.

Paudyal et al. \cite{Parad_Paud} studied the impact of different rendering algorithms on the perceptual quality of light field images. In addition, they investigated the application of 2D objective metrics on light field images. Full-reference and reduced-reference metrics correlate well with the subjective scores. However, current no-reference quality metrics do not correlate well with the subjective scores. In~\cite{RR_QA}, a reduced-reference metric is discussed evaluating the perceptual quality based on light field images and their distorted depth maps. Tamboli et al. \cite{TAMBOLI201642} extend the standard subjective quality evaluation from 2D spatial content to 3D light field displays by proposing a 3D full-reference metric taking into account the spatial-angular nature of the 3D content.  

Adhikarla et al. \cite{Adhikarla879} have conducted subjective studies to evaluate the impact of different distortions on light fields perceived quality during reconstruction, displaying, and transmission. Furthermore, Min et al. \cite{Min2319} proposed a light field quality metric to evaluate quality loss in the case of visual distortions e.g. compression, visualization, and reconstruction. The metric uses global and local spatial information together with angular information for assessing the quality. Ak et al. \cite{Ak6194} examine the existing 2D image quality metrics to analyze their performance on light field content represented as epipolar plane images. Three chosen metrics, NICE \cite{NICE}, GMSD \cite{GMSD}, and MW-PSNR \cite{MW-PSNR}, rely on structural information like gradient and extracted edge information. Tian et al. \cite{Tian387} implemented a full-reference metric for light field images relying on symmetry and depth-based models. This metric explores light field spatial features reflecting color and geometry.

Nevertheless, the methods developed for light field quality assessment are generally experimented on 2D displays which lack depth cues, motion parallax, and aliasing artifacts. Therefore, in this study, our goal is to use light field displays to evaluate \ac{DoF} guided visualization perceptually.

%As mentioned in Section \ref{Background}, light field displays have physical limitation in visualizing 3D content outside display DoF. Therefore, the original content cannot be shown without aliasing artifacts. In addition, applying DoF rendering to mitigate aliasing issues would cause blurring. Therefore, no reference metric is required to evaluate the result of DoF rendering comparing to aliased content. For achieving the mentioned purpose, a subjective study is required that correlates with a proposed no reference model. 

\section{Depth-of-Field Aware Scene Complexity (DASC) Metric}
\label{model}
Estimating the perceived optimal level of blurriness of a light field visualized on a 3D display, requires a model that considers the scene characteristics and correlates with the observers opinion scores, as will be discussed in Section \ref{Subjective Study} and \ref{evaluation}. The characterization of the scene accounts for geometric and position factors of each object, and integrates them into a unified score. To this purpose we use edge density, entropy, and curvature standard deviation as geometric factors. The position factors include the object's size, its overlap with the DoF region, and its position with respect to the DoF region. 

%To estimate the geometric and position factors of each object in the scene, they need to be separated from each other, and this can be done using a segmentation map. Additionally, depth map is essential for computing some geometric and position factors.

The assumption behind the proposed DASC metric is that aliasing distortions are more visible for intricate objects, that are positioned outside of the display's \ac{DoF}. In addition, objects spanning large depth range exhibit varying distortion levels at different depths. Therefore, the perception of \ac{DoF} rendering might be different across individual parts of the object. 
%In this paper, we propose the \ac{DASC} metric that allows to characterize the scene content based on the aforementioned factors.

\subsection{Geometric Factors}
\begin{figure}[h!]
    \centering
    \includegraphics[width=0.95\linewidth]{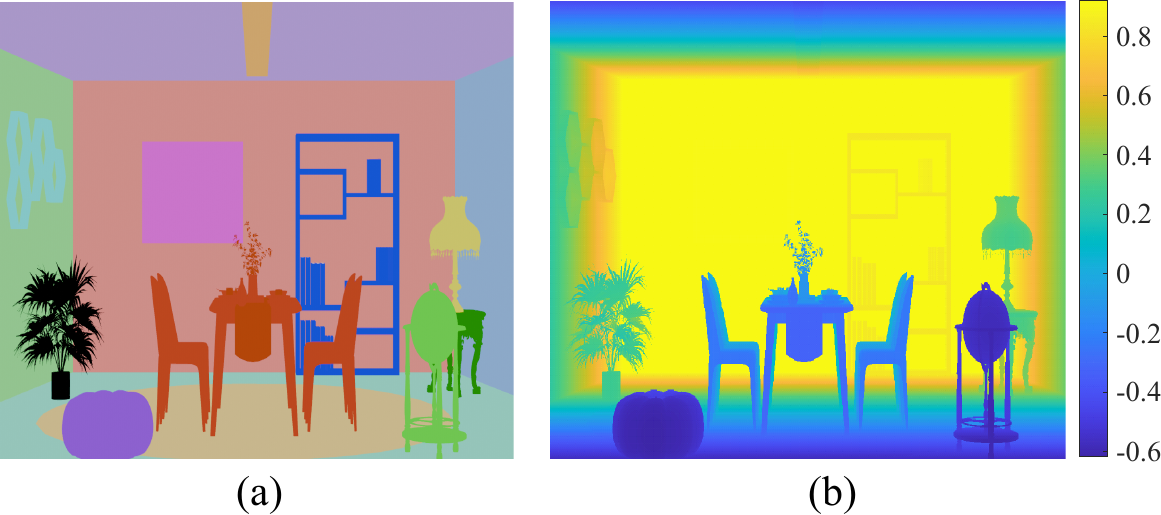}
    \caption{ Example of (a) a segmentation map and (b) depth map when evaluating geometric and position factors. }.
    \label{fig:seg_map}
\end{figure}

The proposed metric evaluates the complexity of each object individually based on three geometric factors: entropy, edge density, and curvature. This requires a segmentation map to identify objects and the depth map for estimating the object's 3D curvature. There are many state-of-the-art methods to generate segmentation maps \cite{SEGNET} \cite{SIS} and depth maps \cite{ren2019deeprobustsingleimage} \cite{khan2021}. In this paper, we utilize Blender \cite{blend} for generating these maps as well as for rendering views in the light field. Figure \ref{fig:seg_map} shows an example of a segmentation and depth map generated from Blender. In the figure, each color in the segmentation map represents an object's area, $\Omega_i$.

\subsubsection{Entropy}
measures the amount of randomness in the color of an object's texture, and shows its complexity which is useful for compression, texture analysis, and segmentation. Entropy within the area of an object, $\Omega_i$ can be evaluated as: \cite{Sparavigna0502},
\begin{gather}
\label{eq:entropy}
\textbf{s}(\textbf{I}(\Omega_i)) = -\sum_{j=0}^{255}p_j(\textbf{I}(\Omega_i)log_2(p_j(\textbf{I}(\Omega_i))),
\end{gather}
where $\textbf{I}(\Omega_i)$ is the color histogram of the object's texture, and $p_j(\textbf{I}(\Omega_i))$ is the probability of $\textbf{I}(\Omega_i)$ for color intensity $j$. High entropy indicates textures with high variations in pixels' color.  

\subsubsection{Edge density} evaluates the number of edge pixels in the area of an object, and can be expressed as \cite{gonzalez2008digital},
\begin{gather}
\label{eq:edge_density}
\textbf{e}(\Omega_i) = \frac{\sum_{\Omega_i}^{}e_{u,v}(\Omega_i)}{|\Omega_i|},
\end{gather}
where $|\Omega_i|$ refers to the number of pixels in the object's area and $e_{u,v}(\Omega_i)$ corresponds to the value of the edge pixel $(u, v)$ within object's area $\Omega_i$  which is derived by edge detection algorithms (Sobel, Canny, or Prewitt) \cite{Gonzalez2017}. High edge density for an object in a scene indicates either complex geometry or texture with intricate patterns. 
\subsubsection{Curvature Standard Deviation} quantifies how much an object's surface deviates from being smooth. To estimate the curvature standard deviation, the first step is to compute the Hessian matrix for each pixel $(u, v)$ inside the object's area $\Omega_i$.
\begin{gather}
\label{eq:Hessian}
h_{u,v}(\Omega_i) = \begin{bmatrix}
        \frac{d^{2}z_{u, v}(\Omega_i)}{d^{2}u} & \frac{d^{2}z_{u, v}(\Omega_i)}{dudv} \\
        \frac{d^{2}z_{u, v}(\Omega_i)}{dudv} & \frac{d^{2}z_{u, v}(\Omega_i)}{d^{2}v}
        \end{bmatrix},
\end{gather}
where $z_{u, v}(\Omega_i)$ is the depth value at the pixel $(u, v)$. The following step is to compute the average of the eigen values of the Hessian matrix which results in the mean curvature $\textbf{C}(\Omega_i)$ with value $c_{u, v}(\Omega_i)$  at pixel coordinate, $(u, v)$.
\begin{gather}
\label{eq:mean_curvature}
c_{u, v}(\Omega_i) = \frac{\lambda^{(1)}_{u, v}(\Omega_i)+\lambda^{(2)}_{u, v}(\Omega_i)}{2}.
\end{gather}
Finally, for an object $\Omega_i$, the standard deviation of the mean curvature is computed as \cite{Hessian_Curvature}:
\begin{gather}
\label{eq:std_curve}
\boldsymbol{\sigma_C}(\Omega_i) = \sqrt{\EX[(\textbf{C}(\Omega_i)-\EX[\textbf{C}(\Omega_i)])^2]}.
\end{gather}
High curvature standard deviation implies greater surface roughness and complexity.

Entropy, edge density, and curvature standard deviation produce outputs with different value ranges. To ensure that each factor has an equal impact, their values should be normalized to range [0, 1]. The normalization is done as,

\begin{gather}
\label{eq:norm_E}
\Bar{\textbf{e}}=\frac{\textbf{e} - \min(\textbf{e})}{\max(\textbf{e}) - \min(\textbf{e})} \\
\label{eq:norm_S}
\Bar{\textbf{s}}=\frac{\textbf{s} - \min(\textbf{s})}{\max\textbf{s}) - \min(\textbf{s})} \\
\label{eq:norm_C}
\boldsymbol{\Bar{\sigma}_C} = \frac{\boldsymbol{\sigma_C} - \min(\boldsymbol{\sigma_C})}{\max(\boldsymbol{\sigma_C}) - \min(\boldsymbol{\sigma_C})},
\end{gather}
where $\Bar{\textbf{e}}$, $\Bar{\textbf{s}}$, and $\boldsymbol{\Bar{\sigma}_C}$ are the normalized edge density, entropy, and curvature standard deviation, respectively. 

\subsection{Position Factors}
The position factors include the object's location relative to the display's \ac{DoF}, object's size, and the extent of its overlap with the \ac{DoF} region. The main assumption for the position factor is that aliasing distortions and blurring can be perceived differently depending on the object's position with respect to the display's screen. 
%and its extension over the depth range.  
\subsubsection{Object's overlap with \ac{DoF}}
The extent of an object's overlaps with the light field display's \ac{DoF} is determined by the number of pixels corresponding to the object's area $\Omega_i$ that lie within the \ac{DoF} region $\Phi$. The ratio can be calculated as, 
\begin{gather}
\label{eq:W_area}
\boldsymbol\omega(\Omega_i)=\frac{|\Omega_i \cap \Phi|}{|\Omega_i|}.
\end{gather}
Higher ratio indicates a lower likelihood of the object being visualized with aliasing distortions.

\subsubsection{\ac{DoF}-aware positioning}
A position factor is evaluated based on the object's location relative to the display's \ac{DoF}. This factor can be computed as,
\begin{gather}
\label{eq:W_min} 
\boldsymbol{d_{\min}}(\Omega_i) = \frac{\min(|\min(\textbf{Z}(\Omega_i)) \pm \frac{d_{\Phi}}{2}|)}{d_\Phi}\\
\label{eq:W_max} 
\boldsymbol{d_{\max}}(\Omega_i) = \frac{\min(|\max(\textbf{Z}(\Omega_i)) \pm \frac{d_{\Phi}}{2}|)}{d_\Phi},
\end{gather}
where $d_{\Phi}$ corresponds to the size of the display's \ac{DoF} - see Figure \ref{fig:feature_size} and Figure \ref{fig:setup}.  The DoF-aware positioning indicates the proximity of the object to the display's \ac{DoF}. Greater value of $\boldsymbol{d_{\max}}(\Omega_i)$ and/or $\boldsymbol{d_{\min}}(\Omega_i)$ indicates that an object spans a long depth range, resulting in varying distortions across its different parts. Scaling the factor with the display’s \ac{DoF} size is motivated by the fact that different \ac{DoF} sizes influence the distortion rate.

\subsubsection{Object's size}
The object size is calculated based on its maximum depth, $\max(\textbf{Z}({\Omega_i}))$, and minimum depth, $\min(\textbf{Z}({\Omega_i}))$, as,
\begin{gather}
\label{eq:length} 
\textbf{l}(\Omega_i)= \frac{|\max(\textbf{Z}(\Omega_i)) - \min(\textbf{Z}_{\Omega_i})|}{d_\Phi}.
\end{gather} This factor represents the size of the object within the display's \ac{DoF} region, and it is also scaled with the display's \ac{DoF} size. Higher values indicate a larger span within the \ac{DoF}, which results in a visualization without aliasing.

\subsection{Aggregation}
Based on the estimated geometric and position factors, the DASC metric  returns a score, $f$, that accounts for  the scene characteristics. It is computed as:

\begin{gather}
\label{eq:aggregation}
f = \frac{1}{m}\sum_{i}^{}(1-\boldsymbol{\omega}(\Omega_i))\boldsymbol\nu(\Omega_i)\boldsymbol{\psi}(\Omega_i),
\end{gather}
where\\
\begin{gather}
\label{eq:aggregation2}
\boldsymbol\nu(\Omega_i) = \frac{\boldsymbol{\Bar{\sigma}_C}(\Omega_i) + \Bar{\textbf{e}}(\Omega_i) + \Bar{\textbf{s}}(\Omega_i)}{3},
\end{gather}

\begin{gather}
\boldsymbol{\psi}(\Omega_i) = \left\{ \begin{array}{cl}
\boldsymbol{d_{\min}}(\Omega_i) & : \textbf{Z}_{\min}(\Omega_i) > \frac{d_{\Phi}}{2}, \space \textbf{Z}_{\max}(\Omega_i) \leq  \frac{d_{\Phi}}{2} \\
 \boldsymbol{d_{\max}}(\Omega_i) & :  \textbf{Z}_{\min}(\Omega_i) \leq  \frac{d_{\Phi}}{2}, \space  \textbf{Z}_{\max}(\Omega_i) > \frac{d_{\Phi}}{2} \\ \boldsymbol\zeta(\Omega_i)
 &  : \  \textbf{Z}_{\min}(\Omega_i), \textbf{Z}_{\max}(\Omega_i) >  \frac{d_{\Phi}}{2}\\
\frac{-\textbf{l}(\Omega_i)}{1-\boldsymbol{\omega}(\Omega_i)} & :  \textbf{Z}_{\min}(\Omega_i), \textbf{Z}_{\max}(\Omega_i)  \leq \frac{d_{\Phi}}{2},
\end{array} \right.
\\
\boldsymbol\zeta(\Omega_i) = \boldsymbol{d_{\min}}(\Omega_i) + \boldsymbol{d_{\max}}(\Omega_i).
\end{gather}
Based on the definition of \ac{DASC} it is relevant to notice that:
\begin{itemize}
  \item \ac{DASC} considers an equal contribution of normalized geometric factors.  Eq. \ref{eq:aggregation2} accounts for this by computing the average of these features.
  \item \ac{DASC} is weighted by the factor $1-\boldsymbol{\omega}(\Omega_i)$ that is the normalized area of the object outside the display's \ac{DoF}. The larger the area outside the display's DoF, the higher the chances that aliasing becomes visible.
  \item If the back or the front part of the object is outside the display's \ac{DoF}, the metric is weighted by $\boldsymbol{d_{\min}}(\Omega_i)$ or $\boldsymbol{d_{\max}}(\Omega_i)$, respectively. A higher weight corresponds to a wider span outside the display's \ac{DoF}, increasing the levels of distortion (e.g., regions closer to the display \ac{DoF} appear less distorted with respect to the farther ones).
  \item If both back and front parts of the object are outside the \ac{DoF}, \ac{DASC} is weighted by $\boldsymbol\zeta(\Omega_i)$ which adds $\boldsymbol{d_{\max}}(\Omega_i)$ and $\boldsymbol{d_{\min}}(\Omega_i)$. Larger weights correspond to higher chances to be visualized with aliasing.
  \item If both back and front parts of the object are inside the \ac{DoF}, \ac{DASC} is weighted by $\frac{-\textbf{l}(\Omega_i)}{1-\boldsymbol{\omega}(\Omega_i)}$ indicating the span of objects that are totally inside the display's \ac{DoF} and visualized without any distortions.
  \item The score value $f$ is scaled by the number of object $m$, which means the metric is independent of the number of objects.
  
\end{itemize}

Overall, \ac{DASC} exploits geometric and position factors to evaluate the scene structure. The value of \ac{DASC} metric ranges from $\left[ -1,  \infty\right)$. In the special case where objects in the scene mostly resides in the display's \ac{DoF}, the value of the \ac{DASC} metric is negative, and bounded between -1 and 0.  The DASC metric will be used to suggest the optimal blurring level that reduces aliasing distortions depending on the content. 
\begin{comment}
 \begin{figure*}[t!]
    \centering
    \includegraphics[width=\textwidth]{images/DepthDist2.pdf}
    \caption{Depth distribution of objects in each scene using a box plot, representing the minimum, maximum, median depth, and the 25th and 75th percentile depths. The shaded region corresponds to the \ac{DoF} of Holovizio 722 RC. }
    \label{fig:ValDepthDist}
\end{figure*}   
\end{comment}

\begin{figure*}[t!]
    \centering
    \includegraphics[width=0.85\textwidth]{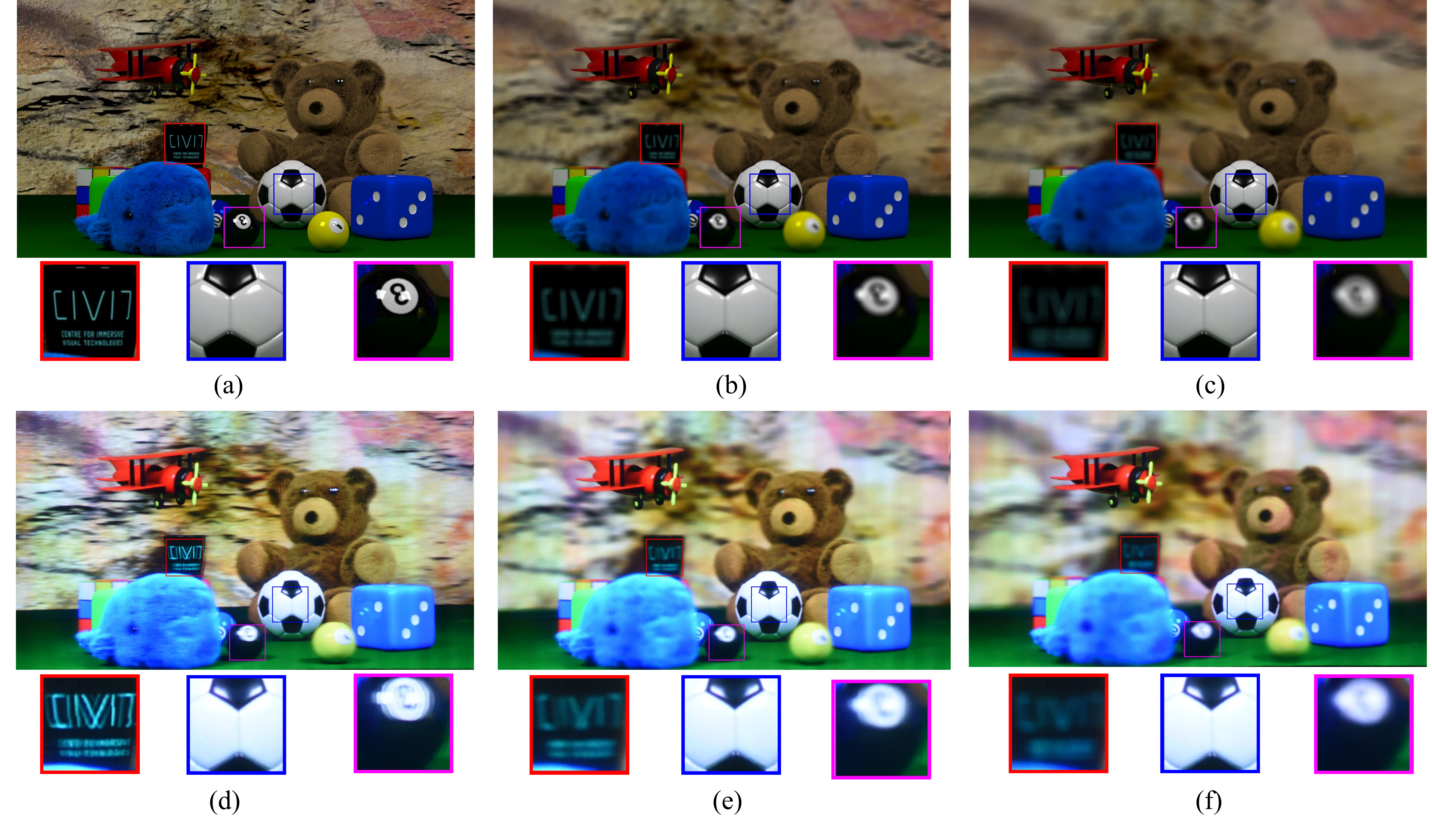}
    \caption{Visualization of different blurring levels $r = \{0, 9, 15\}$ on 2D screen (top row) and Holovizio 722 RC (bottom row). }
    \label{fig:DispVisComparison}
\end{figure*}

\section{Subjective Study}
\label{Subjective Study}
In this section, we describe the conducted subjective studies to understand the preferred level of blurriness for \ac{DoF} guided visualization on light field displays. Initially, we design and create a light field dataset featuring diverse scenes with numerous objects spanning various depths. The designed dataset is visualized on the light field display with various blurring levels. We conduct a subjective test on the prepared dataset to establish a correlation model between the participants' opinion score and the values of the \ac{DASC} metric proposed in the previous section. In the following we will refer to this subjective test as\textit{ correlation study}. Next, we confirm the effectiveness of the correlation model by another subjective experiment with different contents and participants. We will refer to this study as \textit{validation study}. We utilized Holovizio 722 RC as the light field display for the subjective studies \cite{Balogh2008THEHS}.

\subsection{Dataset}
One of the contributions in this paper is the design and creation of a dataset for the subjective studies, which includes nine synthetic light fields rendered by Blender \cite{blend}. According to the circular filtering requirement discussed in \cite{AKbar_antialiasing_LF_display}, the rendered light field has to be full parallax and densely sampled. In the light field dataset, the content within the rendered light field is distributed over a wide depth range, with multiple objects positioned outside the display's \ac{DoF}, making them prone to aliasing distortions. The existing light field datasets \cite{CIVIT_dataset} \cite{Li_2014_CVPR} \cite{Rerabek2016} \cite{dansereau2019liff} do not have the aforementioned features. The central view from \textit{Toys} is shown in Figure \ref{fig:DispVisComparison}a. Other scenes of the dataset are presented in the supplementary material.

%The first six scenes (a-f) are used in the \textit{correlation study} and the last three scenes (g-i) are used in the \textit{validation study}.

%In the dataset, we arranged the content of all scenes in the similar depth range with the aim to have a similar angular and spatial resolutions for the rendered light fields. %Since, \textit{Dragon} and \textit{Camper} contain open space, where the furthest point of the background is significantly distant, we exclude background in these scenes, when computing $b$ to render \ac{DSLF}. This fact decreases the computational complexity, and the impact is minimal and unnoticeable when visualizing these scenes on the light field display.

For the prepared dataset, we analyze the heterogeneity of the dataset by considering different features, in order to avoid results being significantly affected by the scene content. The utilized features are \ac{SI}, \ac{CF}, contrast, \ac{TI}, and depth distribution of scene objects. For detailed information see the supplementary material.

%\ac{SI} computes the pixels standard deviation in each view of the light field after applying Sobel filter \cite{characterization}. \ac{CF} and contrast conveys visual information related to perceptual quality and image naturalness \cite{characterization}. In this study, we use \ac{TI} to estimate the difference between two consecutive views in the light field angular domain \cite{ITURec500}. Figure \ref{fig:features_scenes} shows that  \ac{SI}, \ac{TI}, \ac{CF} and contrast are well-distributed over the span. The depth distribution of each object in the scene provides information about objects placement in the scene with respect to the light field display's \ac{DoF}. Figure \ref{fig:ValDepthDist} illustrates objects' depth distribution for each scene in the dataset. The shaded region is the \ac{DoF} of Holovizio 722 RC. %in the scene where each scene has objects with different depth ranges and positions relative to the display's \ac{DoF}.

Circular filters, mentioned in Section \ref{Background}, are applied to each scene in the dataset to generate different levels of blurriness in the light fields. The aperture radii used in this paper are $r=\{0, 3, 6, 9, 12, 15\}$, where $r=0$ corresponds to sharp, aliased scene, and $r=15$ corresponds to an over-blurred alias-free scene. Figure \ref{fig:DispVisComparison} illustrates the visual difference between multiple blurring levels on the light field display. We use different scenes for the\textit{ correlation study} and \textit{validation study}.

The source code and dataset used in this study are publicly available on \href{https://github.com/kamran-akbar/Perceptual-Assessment-of-Depth-of-Field-Visualization-in-Light-Field-Displays/tree/main}{GitHub\textsuperscript{\textregistered}}. 

% I WOULD SKIP THIS (Robert) The artists whose 3D models are used in the dataset are given appropriately credit in the \href{https://github.com/kamran-akbar/Perceptual-Assessment-of-Depth-of-Field-Visualization-in-Light-Field-Displays/blob/main/Acknowledgments.pdf}{acknowledgment.pdf}. 
\begin{figure*}[t!]
    \centering
    \includegraphics[width=0.8\textwidth]{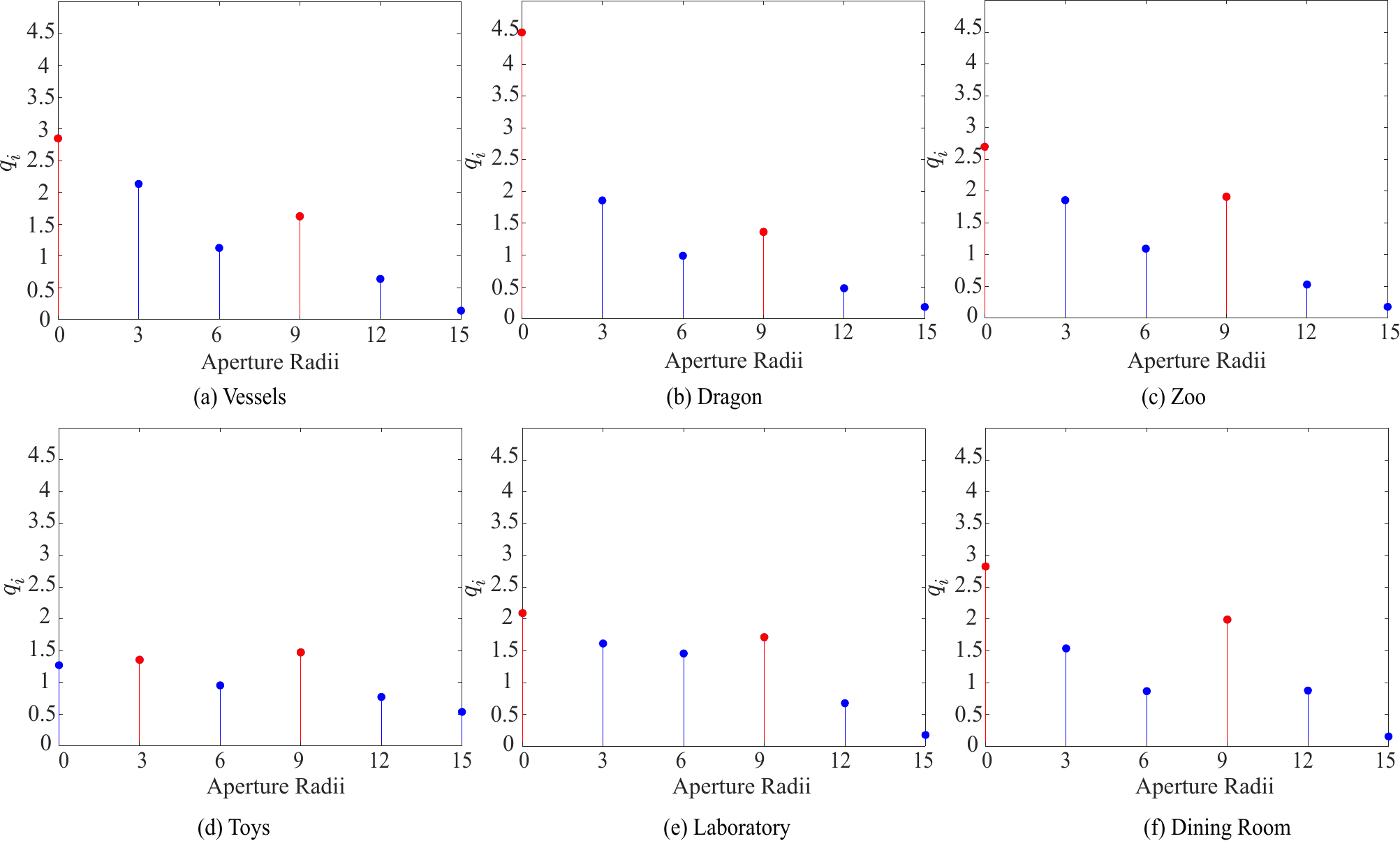}
    \caption{ The result of Bradley-Terry model which demonstrates the preference of different filtering radii for (a) \textit{Vessels}, (b) \textit{Dragon}, (c) \textit{Zoo}, (d) \textit{Toys}, (e) \textit{Laboratory}, and (f) \textit{Dining room}  scenes. The red bars corresponds to the peaks in the opinion scores. }
    \label{fig:bradley}
\end{figure*}
\subsection{Experimental Protocol}
Most of the existing protocols for subjective experiments are designed for 2D displays \cite{ITURec500}. The few studies addressing light field displays \cite{TAMBOLI201642} are not appropriate for our case because the content in the dataset is mainly positioned in the display's \ac{DoF} which does not cause aliasing distortions. Therefore, we established a protocol based on existing standards and recommendations to determine the preference of observers over different levels of blurring. To this aim we selected the double-stimulus impairment scale \cite{ITURec500}. In our setup, each pair contains two light field images with different levels of blurring and are visualized on the light field display. The comparison pairs are chosen based on the square design \cite{Li938} and the total number of comparison for each scene is 9. The aperture radii, which define the blurring amount for each pair of light field images, are selected from $r=\{0,3,6,9,12,15\}$. 
The sequence of the scenes shown to the participants are randomized.

The protocol for both subjective studies is outlined in the following:
\begin{enumerate}
  \item We familiarize the participant with the purpose of the study and the laboratory environment.
  \item We describe the consent form and ask for their signature.
  \item The participants undergo visual acuity (Snellen test) and color blindness tests (Ishihara test) to check their suitability for the participation to the tests. \footnote{We conduct these tests for generality, but they do not impact the study's results as we exclude participants with abnormal vision. }
  \item We introduce participants to the test through a training session. During this phase, participants observe a three sets of light field image pairs with varying blurring amounts, and choose the preferred one. In this way the participants become familiar with the type of distortion and the voting mechanism. The participants opinion score for the tutorial is discarded.
  \item The actual experiment is done by visualizing pairs of light field images on the light field display. The first image is shown to participants for 10 seconds and screen turns into gray for 5 seconds. Then, the second image is shown to the participants for 10 seconds, and the screen turns into gray for 5 seconds. We repeat this procedure for the second time. 
  \item We ask participants to choose their preferred light field image from the visualized pair and move to the next image pair.
\end{enumerate}

\subsubsection{Correlation Study} The purpose of the correlation study is to establish a correlation model between the participants opinion scores and \ac{DASC}. %For this study, we have done a pilot study to verify the effectiveness of the designed protocol. The number of pilot participants is four. 
 31 participants carried out the test and, among these, 4 were discarded due to abnormal vision or for not having followed the protocol. The duration of the test is 70-75 minutes with 10-15 minutes break after the first 30 minutes. The participants' mean age is 29 $\pm$ 4.9 years. The participants group includes $42\%$ females and $58\%$ males. The correlation study consists of 6 different scenes and in total 54 comparison were done by each participant.
\subsubsection{Validation Study}
The purpose of the validation study is to confirm the correctness of the correlation model. The experiment is done with 14 participants. The duration of the validation study is 40-45 minutes. The participants' mean age was 25 $\pm$ 6.15 years. The participants group includes $43\%$ females and $57\%$ males. The validation study consists of 3 different scenes and in total 27 comparison were done by each participant.

\section{Evaluation}
\label{evaluation}
In this section, first, we analyze the result of the correlation study, and build a model that correlates the participants opinion scores with the \ac{DASC} metric proposed in Section \ref{model}. Thereafter, we examine the result of the validation study to confirm the effectiveness of the correlation model.

\subsection{Correlation Study Analysis}
\label{Analysis}

We utilized Bradley-Terry model to analyze participants preferences. Bradley-Terry is a model that estimates the probability of preferring stimulus $i$ over stimulus $j$ as
\begin{gather}
\label{eq:preference}
Pr(r_i > r_j) = \frac{q_i}{q_i + q_j},
\end{gather}
where $q_i$ is the stimulus $i$ score and $q_i \geq 0,\space q_i \in \mathbb{R}$. The value of $q_i$ is evaluated as \cite{BradleyTerry}
\begin{gather}
\label{eq:score1}
q_i^k = \frac{\sum_{j}^{}w_{ij}}{\sum_{j}^{}\frac{(w_{ij}+w_{ji})}{(q_i^{k-1}+q_j^{k-1})}},
\end{gather}
where $w_{ij}$ is the number of times when stimulus $i$ is preferred over stimulus $j$. Since there is no close-form solution for Equation \ref{eq:score1}, it is solved iteratively. $q^k_i$ is the stimulus $i$ score after the $k^\text{th}$ iteration. $q^k_i$ is normalized as: 
%Alternatively, Newman et al.~\cite{Newman1086} proposed an approach for fast convergence of $q_i$ as:
\begin{comment}
\begin{gather}
\label{eq:score2}
q^k_i = \frac{\sum_{j}^{}\frac{w_{ij}q^{k-1}_j}{q_i^{k-1}+q_j^{k-1}}}{\sum_{j}^{}\frac{w_{ji}}{(q_i^{k-1}+q_j^{k-1})}},
\end{gather}    
\end{comment}

\begin{gather}
\label{eq:normalize_P}
q^{(k+1)}_i = \frac{q^k_i}{(\prod_{j}^{}q^k_j)^{\frac{1}{n}}}.
\end{gather}
In this study, we use Equation \ref{eq:score1} to estimate participants opinion score, $q_i$ for all aperture radii.
Figure \ref{fig:bradley} illustrates the opinion scores for each scene utilized in the correlation study. As seen in the figure, it exhibits two peaks, marked in red. This suggests two possible explanations for the observed peaks in opinion scores. First, opinion scores could be observer-based. This means that there are two distinct groups of participants whose preference is consistent regardless of the content. The second explanation is that preferences are content-based. This corresponds to the change of participants' preferred blurring level depending on the content.

\begin{figure}[h!]
    \centering
    \includegraphics[width=0.9\linewidth]{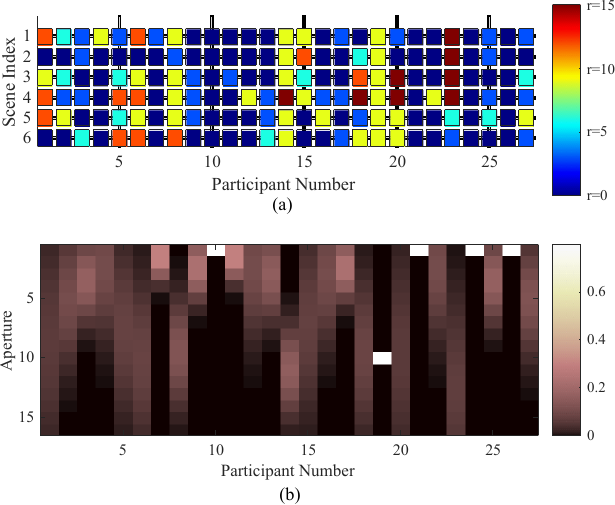}
    \caption{ (a) The preferred aperture radii $r$ for each participant and each scene.  (b) The normal distribution of participants preference for aperture radii across all scenes. }
    \label{fig:separateBTAnalysis}
\end{figure}

To understand the reason behind two peaks, we reevaluate the Bradley Terry model for each participant separately. Figure \ref{fig:separateBTAnalysis} (a) illustrates the preferred aperture radii for each participant and each scene. Thereafter, for each participant, we compute the mean and standard deviation of preferred aperture radii among all scenes, and estimate the normal distribution as illustrated in Figure \ref{fig:separateBTAnalysis} (b). If a participant's preferred blurring level does not change regardless of the scene, its standard deviation is zero and there is a dominant peak in the mean value of the distribution e.g. participant 10 and 19 in Figure \ref{fig:separateBTAnalysis}. As seen in Figure \ref{fig:separateBTAnalysis}, only 5 out of 27 participants, $18.52\%$, have consistent preference among different scenes. The preference of the remaining participants depends on the content. Therefore, we can infer that the observation of two peaks in Figure \ref{fig:bradley} is content-based.

As shown in Figure \ref{fig:bradley}, the difference of two peaks in the opinion scores are low for some scenes, e.g. \textit{Toys} and \textit{Laboratory}. Therefore, we statistically assess the similarity between aperture radius pairs corresponding to the two peaks using a \ac{LRT}\cite{LRT}. In this test, the null hypothesis corresponds to the case that two compared aperture radii are equally preferred, and the alternative hypothesis states the corresponding aperture radii are differently perceived. \ac{LRT} follows $\chi^2$ distribution, and we perform $\chi^2$ test to accept or reject the null hypothesis \cite{Wilk_Theorm}. In the similarity check, we use $\chi^2_{df=1}(\alpha=0.05) = 3.84$. The degrees of freedom, $df=1$, as the comparison is made between pairs of aperture radii. A significance level of $\alpha=0.05$ is adopted, as it is commonly used in the state-of-the-art literature. \ac{LRT} is computed for each scene as,
\begin{gather}
\label{eq:LRT}
\lambda = -2\log(\frac{L(r_i,r_j;\theta_0)}{ L(r_i,r_j;\theta_1)}), \space \space r_i\neq r_j, 
\end{gather}
where $\theta_0$ and $\theta_1$ correspond to null and alternative hypotheses, respectively. If $\lambda < \chi^2_{df=1}(\alpha=0.05)$, the null hypothesis is accepted. In the following section we will describe how we utilize \ac{LRT} to approximate the preferred blurring level for each scene and correlates it with the \ac{DASC} metric value.

\begin{figure*}[t!]
    \centering
    \includegraphics[width=0.8\textwidth]{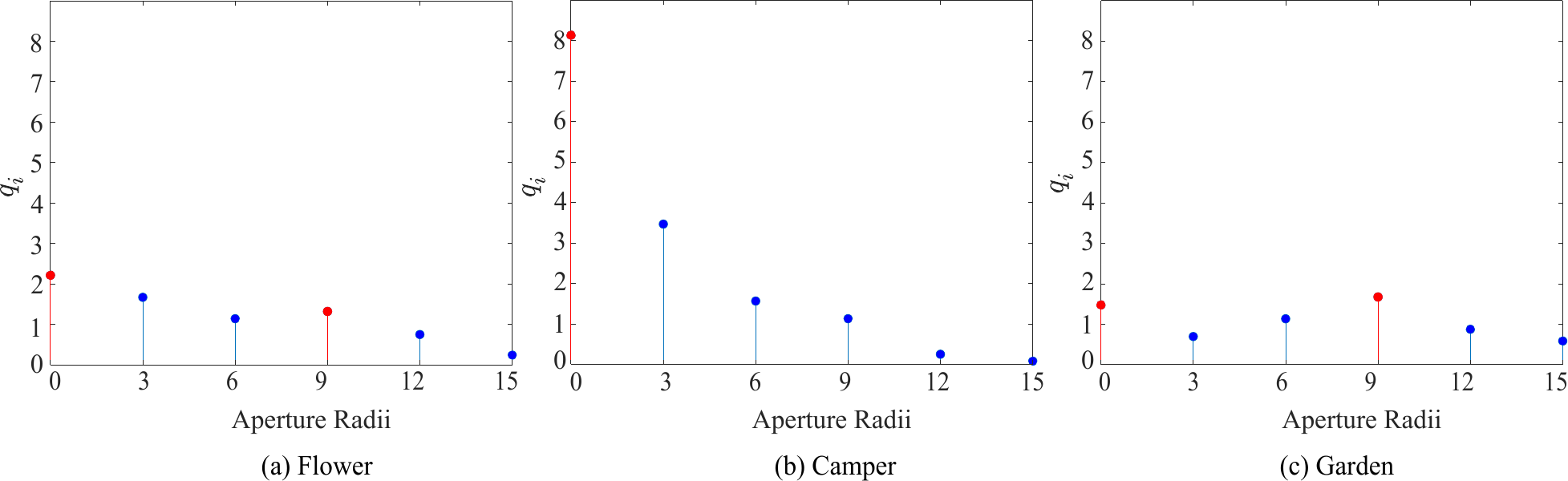}
    \caption{The result of Bradley-Terry model which demonstrates the preference of different filtering radii for validation scenes, (a) \textit{Flower}, (b) \textit{Camper}, and (c) \textit{Garden}. }
    \label{fig:BTval}
\end{figure*}
\begin{table*}[t!]
\caption{The similarity check between aperture radii pairs using \ac{LRT} $\lambda$ and $\chi^2$ test $(df=1, \alpha=0.05)$.}
\label{tab:LRT_Chi_2}
\centering
\begin{tblr}{
  cell{1}{1} = {c=2,r=2}{},
  cell{1}{3} = {c=6}{c},
  cell{1}{9} = {c=3}{},
  cell{3}{1} = {r=2}{},
  vlines,
  hline{1,3,5} = {-}{},
  hline{2} = {3-11}{},
  hline{4} = {2-11}{},
}
           &        & \textbf{Correlation} &        &      &      &            &             & ~ ~ ~ ~ ~ ~\textbf{Validation} &        &        \\
           &        & \textit{Vessels}     & \textit{Dragon} & \textit{Zoo}  & \textit{Toys} & \textit{Laboratory} & \textit{Dining Room} & \textit{Flower}                & \textit{Camper} & \textit{Garden} \\
\textbf{Radii Pair} & (0, 9) & 4.08        & 8.81   & \textbf{1.61} & \textbf{0.33} & \textbf{0.73}       & \textbf{2.33}        & \textbf{2.17}                  & 7.77   & \textbf{0.28}   \\
           & (0, 3)  & 0.04           & 8.15      & 0.18    & 0.04 & \textbf{0.04}          & 4.23           & 0.0                     & 2.64      & 1.07      
\end{tblr}
\end{table*}
\subsection{Correlation Model}
\label{corr}
In this section, we establish a model that correlates the \ac{DASC} metric value for all scenes with the opinion scores from the correlation study. To build the correlation model, first, we estimate the preferred aperture radius for each scene used in the correlation study. As shown in Figure \ref{fig:bradley}, we pick aperture radii that correspond to the two peaks in the opinion score, and check their similarity using \ac{LRT} values provided in Table \ref{tab:LRT_Chi_2}. If null hypothesis is rejected, the two aperture radii are significantly different, we select the one with higher opinion score value. Otherwise, both choices are similarly preferred. In our approach, we pick the aperture radius that is close and greater than the optimal radius $\hat{r}$ mentioned in Table I. With this assumption, we eliminate the aliased content, and display it with smooth motion parallax. 

There are mainly two reasons that null hypothesis is accepted. First, the visual difference between two consecutive aperture radii e.g. $(0, 3)$ is negligible and participants do not perceive the difference. Second, depending on the content,  participants may show a similar preference for both sharp, aliased scenes and blurred, anti-aliased 3D content. The main objective of the correlation model is to estimate the range of \ac{DASC} metric values for which this behavior occurs. The bolded values in Table \ref{tab:LRT_Chi_2} indicate instances where the null hypothesis is accepted and are used to approximate the preferred level of blurring for each scene.

In all scenes, except \textit{Toys}, the two aperture radii peaks are $r_1=0$ and $r_2=9$; for the \textit{Toys} scene, the peaks are $r_1=3$ and $r_2=9$. For the \textit{Vessels} and \textit{Dragon} scenes, the two peaks differ significantly as seen in Table \ref{tab:LRT_Chi_2} and $r=0$ is selected as the preferred aperture radius because it has higher opinion score value. In contrast, for \textit{Zoo}, \textit{Laboratory}, and \textit{Dining Room}, the difference between the two peaks is negligible. Therefore, we choose $r_2=9$ as the preferred aperture radius since $r_2>\hat{r}\gg r_1$. Furthermore, for \textit{Toys} scene, there is no direct comparison available between $r_1=3$ and $r_2=9$. Since in Table  \ref{tab:LRT_Chi_2}, the \ac{LRT} for the aperture radii pair $(0,3)$ is approximately zero, we can conclude that both aperture radii have equal preference. To address this, we use $r_3=0$ and perform the similarity check between $r_3$, $r_1$ and $r_2$, see Table \ref{tab:LRT_Chi_2}. Therefore, we use aperture radius $r_3=0$ instead of $r_1=3$ to compare against
$r_2=9$, and there is negligible difference among them. Hence $r_2>\hat{r}$, we conclude that $r_2=9$ is the preferred aperture radius for the \textit{Toys} scene.

\begin{figure}[h!]
    \centering
    \includegraphics[width=0.7\linewidth]{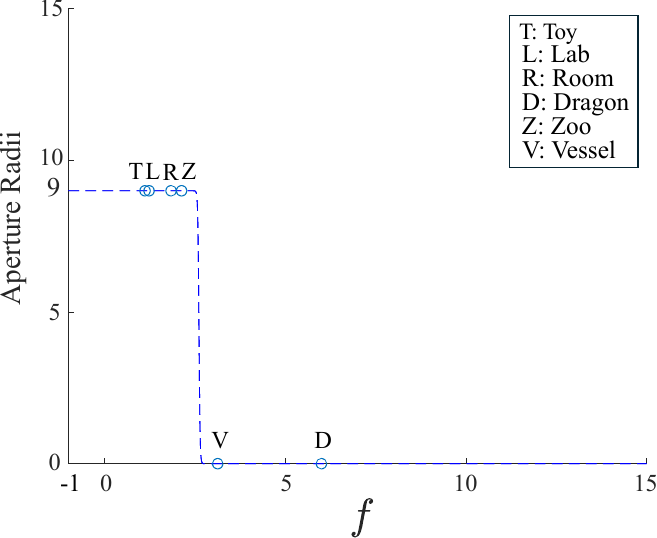}
    \caption{ The correlation between the chosen circular filter radii and the proposed model value for experimented scenes. }
    \label{fig:correlation}
\end{figure}

Following the above explanation, the preferred blurring level for scenes, \textit{Vessels} and \textit{Dragon} is $r=0$, and for scenes \textit{Toys}, \textit{Laboratory}, \textit{Dining Room}, and \textit{Zoo} is $r=9$. Since, there are two types of preferences, we used a non-linear monotonic sigmoid function \cite{Goodfellow-et-al-2016} to correlate the experiment results with the value of the \ac{DASC} metric, $f$. The sigmoid function utilized for fitting is:
\begin{gather}
\label{eq:sigmoid}
    r(f) = \frac{\kappa}{1+e^{\beta(f-\gamma)}},
\end{gather}
where $\kappa$, $\beta$, and $\gamma$ are function coefficients, and their values are determined by MATLAB\textsuperscript{\textregistered} \cite{matlab} non-linear model fitting function, \textit{fitnlm()}. The obtained values are $\kappa=21.9$, $\beta=4.5$, and $\gamma=9.0$. The correlation model is visualized in Figure \ref{fig:correlation}.
According to the model, the estimated aperture radius for scenes with $f$ value between $\left[-1,\space0\right]$, is $r=9$. Scenes with content fully positioned in the display's \ac{DoF} (i.e. no aliasing artifact) have $f$ value within this range. However, filtering scenes without aliasing does not impact their appearance on the display.

\subsection{Validation Study Analysis}
\begin{figure}[h!]
    \centering
    \includegraphics[width=0.7\linewidth]{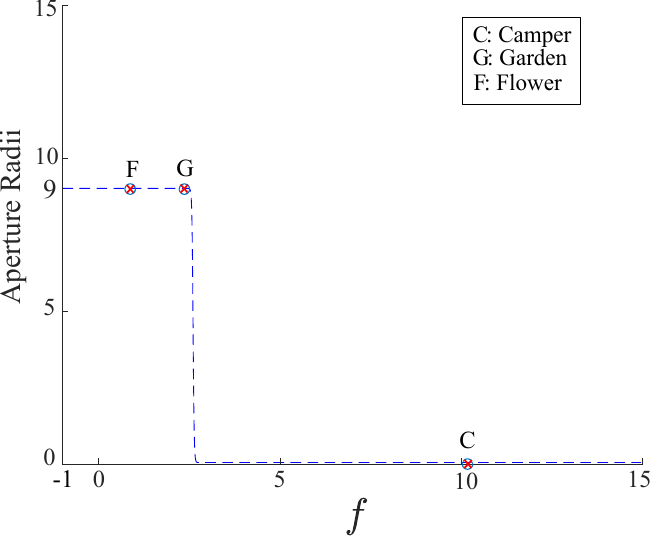}
    \caption{ The comparison of the proposed model's value between preferred aperture radii from validation study and the predicted preferred radii, based on the correlation function. }
    \label{fig:valcorrelation}
\end{figure}

As mentioned in Section \ref{Subjective Study}, the validation study is done separately with a new set of participants and scenes visualized on the light field display. The primary goal of the validation study is to confirm the correlation model's prediction between the DASC metric and the preferred aperture radii. In the validation study, we estimate the participants opinion scores using Equation \ref{eq:score1} as shown in Figure \ref{fig:BTval}. Furthermore, we perform a similarity check between pairs of aperture radii used in the validation study, with the results presented in Table \ref{tab:LRT_Chi_2}. Thereafter, we use this information as discussed in Section~\ref{corr}, to estimate the preferred blurring level for each scene in the validation study. Thus, for scenes \textit{Garden}, \textit{Camper}, and \textit{Flower}, the preferred aperture radius is $9$, $0$, and $9$, respectively. As shown in Figure \ref{fig:valcorrelation}, for these scenes, the estimated blurring levels obtained from the correlation model align perfectly with those from the validation study.

%According to the validation study, the preferred aperture radius for \textit{Flower}, \textit{Garden}, and \textit{Camper} are $r=9$, $r=12$, and $r=0$, respectively. In the next step, we used DASC to estimate the preferred blurring level for these scene. The predicted values are $r=9$, $r=9$, $r=0$, correspondingly. The predicted values for \textit{Camper} and \textit{Flower} matched with the results of validation study. However, for \textit{Garden}, participants prefer more blurred version of the scene to observe. However, with respect to Figure \ref{fig:FSPrefVal} (b), the \ac{PMF} summation of occurrences for which the aperture radius, $r=9$, is selected as a first or second choice, is high enough to satisfy the participants.  
%the participants scores corresponding to each scene is visualized in Figure. The analysis shows that the Camper scene is preferred to be seen without blurring. Respecting the Garden and Flower scenes, participants are divided into two groups which one group prefers sharp and the other prefers to see the content without aliasing. For Flower scene, the second preference of $83\%$ participants who select $r=0$ as their main preference is $r=9$. Also, for the Garden scene, the corresponding percentage is $50\%$. The computed values for Flower, Camper, and Garden scenes respecting the proposed model in Section \ref{model} is shown in Figure \ref{fig:valcorrelation}.

\section{Conclusion}
Light field displays have limited angular and spatial resolutions, thus, the display's \ac{DoF} is limited and aliasing distortions become visible for content visualized outside the \ac{DoF} area. In this study, we have conducted \ac{DoF} guided visualization assessment on light field display to investigate perceptual preference over different blurring levels. To achieve this, we proposed a metric, denoted as \ac{DASC}, that quantifies the scene complexity based on  geometric and position factors of the objects within the scene. Moreover, we have designed and created a dataset containing numerous objects at different depths and we rendered it with different levels of blurring. We developed a protocol for conducting a subjective experiment on light field displays to realize the observers' preference over different levels of blurring. Ultimately, using the estimated opinion scores from the correlation study, we investigated the correlation between the preferences expressed by the participants and the \ac{DASC} metric, and we verified its correctness through a validation study. The proposed model efficiently predicts the preferred level of blurring for any given 3D scene.

While the proposed model based on the DASC metric effectively predicts the required blurring of a 3D scene for optimal visualization on a given 3D display, several aspects could be further refined to enhance the results presented in this paper. First, it would be beneficial to perform a more extensive subjective study with a larger number of scenes and more participants. The main challenges here lie in the preparation of the light fields for different scenes, which is very time consuming, and the duration of the subjective experiment itself, which increases linearly with the number of scenes. Second, the model could be validated across various light field displays. The difficulty here lies in the limited availability of such displays. Third, it would be useful, when generating new scenes, to have them distributed uniformly with respect to the value $f$ of the DASC metric. Although the current selection of scenes has a relatively balanced distribution, there are missing cases for extreme values of $f$. Fourth, incorporating light field saliency has the potential to further improve the proposed model. This is particularly pertinent due to the high complexity (number of objects and occlusions) we have in our scenes. However, the challenge in integrating saliency arises from the lack of existing research that considers light field saliency within the framework used in this study and as such that would require separate investigation.

%OLD VERION *** Since this problem is addressed for the first time, as a future work, we can use faster \ac{DoF} rendering methods than circular filtering used in this paper since it requires a \ac{FP} \ac{DSLF}, and its preparation is time-consuming. Moreover, the duration of subjective experiments highly depends on the number of scenes. Therefore, the number of scenes should be sufficient in a way that does not prolong the experiment and produce reliable result. As a future work, we can benefit from a larger dataset and more participants to prove the validity of our model. Moreover, we can examine the uniformity of scenes' distributions in the \ac{DASC} metric to estimate the transition region of our proposed model, more accurately. We conducted our study, only on one light field display with a distinct \ac{DoF}. As a continuation of our work, the result of our study can be confirmed by conducting the same experiment on several light field displays with different \ac{DoF}. Another potential future direction is using light field saliency to improve DASC model and give higher weights to more salient objects. ***

\bibliographystyle{IEEEtran}
\bibliography{Journal}

\begin{thebibliography}{10}
\providecommand{\url}[1]{#1}
\csname url@samestyle\endcsname
\providecommand{\newblock}{\relax}
\providecommand{\bibinfo}[2]{#2}
\providecommand{\BIBentrySTDinterwordspacing}{\spaceskip=0pt\relax}
\providecommand{\BIBentryALTinterwordstretchfactor}{4}
\providecommand{\BIBentryALTinterwordspacing}{\spaceskip=\fontdimen2\font plus
\BIBentryALTinterwordstretchfactor\fontdimen3\font minus \fontdimen4\font\relax}
\providecommand{\BIBforeignlanguage}[2]{{%
\expandafter\ifx\csname l@#1\endcsname\relax
\typeout{** WARNING: IEEEtran.bst: No hyphenation pattern has been}%
\typeout{** loaded for the language `#1'. Using the pattern for}%
\typeout{** the default language instead.}%
\else
\language=\csname l@#1\endcsname
\fi
#2}}
\providecommand{\BIBdecl}{\relax}
\BIBdecl

\bibitem{levoy_LightField}
\BIBentryALTinterwordspacing
M.~Levoy and P.~Hanrahan, \emph{Light Field Rendering}, 1st~ed.\hskip 1em plus 0.5em minus 0.4em\relax New York, NY, USA: Association for Computing Machinery, 2023. [Online]. Available: \url{https://doi.org/10.1145/3596711.3596759}
\BIBentrySTDinterwordspacing

\bibitem{bregovic_signal_2019}
R.~Bregovic, E.~Sahin, S.~Vagharshakyan, and A.~Gotchev, \emph{Signal Processing Methods for Light Field Displays}.\hskip 1em plus 0.5em minus 0.4em\relax Springer.

\bibitem{AKbar_antialiasing_LF_display}
K.~Akbar and R.~Bregovic, ``Antialiasing filtering for projection-based light field displays,'' in \emph{2023 International Symposium on Image and Signal Processing and Analysis (ISPA)}, 2023, pp. 1--6.

\bibitem{Balogh2008THEHS}
\BIBentryALTinterwordspacing
T.~Balogh, P.~T. Kov{\'a}cs, Z.~Dobranyi, A.~Barsi, Z.~Megyesi, Z.~Ga{\'a}l, and G.~Balogh, ``The holovizio system - new opportunity offered by 3d displays,'' 2008. [Online]. Available: \url{https://api.semanticscholar.org/CorpusID:110209951}
\BIBentrySTDinterwordspacing

\bibitem{Zwicker}
M.~Zwicker, W.~Matusik, F.~Durand, and H.~Pfister, ``Antialiasing for automultiscopic 3d displays,'' in \emph{Proceedings of the 17th Eurographics Conference on Rendering Techniques}, ser. EGSR '06.\hskip 1em plus 0.5em minus 0.4em\relax Goslar, DEU: Eurographics Association, 2006, p. 73–82.

\bibitem{Isaksen}
\BIBentryALTinterwordspacing
A.~Isaksen, L.~McMillan, and S.~J. Gortler, ``Dynamically reparameterized light fields,'' ser. SIGGRAPH '00.\hskip 1em plus 0.5em minus 0.4em\relax USA: ACM Press/Addison-Wesley Publishing Co., 2000, p. 297–306. [Online]. Available: \url{https://doi.org/10.1145/344779.344929}
\BIBentrySTDinterwordspacing

\bibitem{Kamran4738}
\BIBentryALTinterwordspacing
K.~Akbar and R.~Bregovic, ``Depth-of-field guided rendering for light field displays,'' \emph{Electronic Imaging}, vol.~36, no.~10, pp. 367--1--367--1, 2024. [Online]. Available: \url{https://library.imaging.org/ei/articles/36/10/IPAS-367}
\BIBentrySTDinterwordspacing

\bibitem{Wetzstein}
\BIBentryALTinterwordspacing
G.~Wetzstein, D.~Lanman, M.~Hirsch, and R.~Raskar, ``Tensor displays: compressive light field synthesis using multilayer displays with directional backlighting,'' vol.~31, no.~4, Jul. 2012. [Online]. Available: \url{https://doi.org/10.1145/2185520.2185576}
\BIBentrySTDinterwordspacing

\bibitem{Kovacs}
P.~T. Kovács, R.~Bregović, A.~Boev, A.~Barsi, and A.~Gotchev, ``Quantifying spatial and angular resolution of light-field 3-d displays,'' \emph{IEEE Journal of Selected Topics in Signal Processing}, vol.~11, no.~7, pp. 1213--1222, 2017.

\bibitem{singleImage}
K.~Akbar and R.~Bregovic, ``Single-image evaluation of angular and spatial resolution for projection-based light field displays,'' in \emph{2023 IEEE 25th International Workshop on Multimedia Signal Processing (MMSP)}, 2023, pp. 1--6.

\bibitem{electronics_Kara_0953}
\BIBentryALTinterwordspacing
P.~A. Kara, R.~R. Tamboli, E.~Shafiee, M.~G. Martini, A.~Simon, and M.~Guindy, ``Beyond perceptual thresholds and personal preference: Towards novel research questions and methodologies of quality of experience studies on light field visualization,'' \emph{Electronics}, vol.~11, no.~6, 2022. [Online]. Available: \url{https://www.mdpi.com/2079-9292/11/6/953}
\BIBentrySTDinterwordspacing

\bibitem{Visual_Perception}
\BIBentryALTinterwordspacing
A.~Gill, E.~Zerman, C.~Ozcinar, and A.~Smolic, ``A study on visual perception of light field content,'' \emph{CoRR}, vol. abs/2008.03195, 2020. [Online]. Available: \url{https://arxiv.org/abs/2008.03195}
\BIBentrySTDinterwordspacing

\bibitem{Vis_Technique}
F.~Battisti, M.~Carli, and P.~L. Callet, ``A study on the impact of visualization techniques on light field perception,'' in \emph{2018 26th European Signal Processing Conference (EUSIPCO)}, 2018, pp. 2155--2159.

\bibitem{characterization}
P.~Paudyal, J.~Gutiérrez, P.~Le~Callet, M.~Carli, and F.~Battisti, ``Characterization and selection of light field content for perceptual assessment,'' in \emph{2017 Ninth International Conference on Quality of Multimedia Experience (QoMEX)}, 2017, pp. 1--6.

\bibitem{Parad_Paud}
P.~Paudyal, F.~Battisti, P.~Le~Callet, J.~Gutiérrez, and M.~Carli, ``Perceptual quality of light field images and impact of visualization techniques,'' \emph{IEEE Transactions on Broadcasting}, vol.~67, no.~2, pp. 395--408, 2021.

\bibitem{RR_QA}
P.~Paudyal, F.~Battisti, and M.~Carli, ``Reduced reference quality assessment of light field images,'' \emph{IEEE Transactions on Broadcasting}, vol.~65, no.~1, pp. 152--165, 2019.

\bibitem{TAMBOLI201642}
\BIBentryALTinterwordspacing
R.~R. Tamboli, B.~Appina, S.~Channappayya, and S.~Jana, ``Super-multiview content with high angular resolution: 3d quality assessment on horizontal-parallax lightfield display,'' \emph{Signal Processing: Image Communication}, vol.~47, pp. 42--55, 2016. [Online]. Available: \url{https://www.sciencedirect.com/science/article/pii/S0923596516300674}
\BIBentrySTDinterwordspacing

\bibitem{Adhikarla879}
V.~K. Adhikarla, M.~Vinkler, D.~Sumin, R.~K. Mantiuk, K.~Myszkowski, H.-P. Seidel, and P.~Didyk, ``Towards a quality metric for dense light fields,'' in \emph{2017 IEEE Conference on Computer Vision and Pattern Recognition (CVPR)}, 2017, pp. 3720--3729.

\bibitem{Min2319}
X.~Min, J.~Zhou, G.~Zhai, P.~Le~Callet, X.~Yang, and X.~Guan, ``A metric for light field reconstruction, compression, and display quality evaluation,'' \emph{IEEE Transactions on Image Processing}, vol.~29, pp. 3790--3804, 2020.

\bibitem{Ak6194}
A.~Ak and P.~Le-Callet, ``Investigating epipolar plane image representations for objective quality evaluation of light field images,'' in \emph{2019 8th European Workshop on Visual Information Processing (EUVIP)}, 2019, pp. 135--139.

\bibitem{NICE}
D.~M. Rouse and S.~S. Hemami, ``Natural image utility assessment using image contours,'' in \emph{Proceedings of the 16th IEEE International Conference on Image Processing}, ser. ICIP'09.\hskip 1em plus 0.5em minus 0.4em\relax IEEE Press, 2009, p. 2193–2196.

\bibitem{GMSD}
W.~Xue, L.~Zhang, X.~Mou, and A.~C. Bovik, ``Gradient magnitude similarity deviation: A highly efficient perceptual image quality index,'' \emph{IEEE Transactions on Image Processing}, vol.~23, no.~2, pp. 684--695, 2014.

\bibitem{MW-PSNR}
D.~Sandi{\'c}-Stankovi{\'c}, D.~Kukolj, and P.~Le~Callet, ``Dibr-synthesized image quality assessment based on morphological multi-scale approach,'' \emph{EURASIP Journal on Image and Video Processing}, vol. 2017, no.~1, p.~4, 2016.

\bibitem{Tian387}
Y.~Tian, H.~Zeng, J.~Hou, J.~Chen, J.~Zhu, and K.-K. Ma, ``A light field image quality assessment model based on symmetry and depth features,'' \emph{IEEE Transactions on Circuits and Systems for Video Technology}, vol.~31, no.~5, pp. 2046--2050, 2021.

\bibitem{SEGNET}
V.~Badrinarayanan, A.~Kendall, and R.~Cipolla, ``Segnet: A deep convolutional encoder-decoder architecture for image segmentation,'' \emph{IEEE Transactions on Pattern Analysis and Machine Intelligence}, vol.~39, no.~12, pp. 2481--2495, 2017.

\bibitem{SIS}
\BIBentryALTinterwordspacing
G.~Csurka, R.~Volpi, and B.~Chidlovskii, ``Semantic image segmentation: Two decades of research,'' 2023. [Online]. Available: \url{https://arxiv.org/abs/2302.06378}
\BIBentrySTDinterwordspacing

\bibitem{ren2019deeprobustsingleimage}
\BIBentryALTinterwordspacing
H.~Ren, M.~El-khamy, and J.~Lee, ``Deep robust single image depth estimation neural network using scene understanding,'' 2019. [Online]. Available: \url{https://arxiv.org/abs/1906.03279}
\BIBentrySTDinterwordspacing

\bibitem{khan2021}
N.~Khan, M.~H. Kim, and J.~Tompkin, ``Edge-aware bidirectional diffusion for dense depth estimation from light fields,'' in \emph{British Machine Vision Conference (BMVC)}, 2021.

\bibitem{blend}
\BIBentryALTinterwordspacing
B.~O. Community, \emph{Blender - a 3D modelling and rendering package}, Blender Foundation, Stichting Blender Foundation, Amsterdam, 2018. [Online]. Available: \url{http://www.blender.org}
\BIBentrySTDinterwordspacing

\bibitem{Sparavigna0502}
\BIBentryALTinterwordspacing
A.~C. Sparavigna, ``Entropy in image analysis,'' \emph{Entropy}, vol.~21, no.~5, 2019. [Online]. Available: \url{https://www.mdpi.com/1099-4300/21/5/502}
\BIBentrySTDinterwordspacing

\bibitem{gonzalez2008digital}
\BIBentryALTinterwordspacing
R.~C. Gonzalez and R.~E. Woods, \emph{Digital image processing}.\hskip 1em plus 0.5em minus 0.4em\relax Upper Saddle River, N.J.: Prentice Hall, 2008. [Online]. Available: \url{http://www.amazon.com/Digital-Image-Processing-3rd-Edition/dp/013168728X}
\BIBentrySTDinterwordspacing

\bibitem{Gonzalez2017}
\BIBentryALTinterwordspacing
C.~I. Gonzalez, P.~Melin, J.~R. Castro, and O.~Castillo, \emph{Edge Detection Methods and Filters Used on Digital Image Processing}.\hskip 1em plus 0.5em minus 0.4em\relax Cham: Springer International Publishing, 2017, pp. 11--16. [Online]. Available: \url{https://doi.org/10.1007/978-3-319-53994-2_3}
\BIBentrySTDinterwordspacing

\bibitem{Hessian_Curvature}
A.~Bhattacharya and J.~Wall, ``Hessian estimates for the lagrangian mean curvature flow,'' \emph{Calculus of Variations and Partial Differential Equations}, vol.~63, 08 2024.

\bibitem{CIVIT_dataset}
S.~Moreschini, F.~Gama, R.~Bregovic, and A.~Gotchev, ``\BIBforeignlanguage{English}{{CIVIT} dataset: Horizontal-parallax-only densely-sampled light-fields},'' in \emph{\BIBforeignlanguage{English}{European Light Field Imaging (ELFI) Workshop}}, 2019, european Light Field Imaging Workshop ; Conference date: 04-06-2019 Through 06-06-2019.

\bibitem{Li_2014_CVPR}
N.~Li, J.~Ye, Y.~Ji, H.~Ling, and J.~Yu, ``Saliency detection on light field,'' in \emph{The IEEE Conference on Computer Vision and Pattern Recognition (CVPR)}, June 2014.

\bibitem{Rerabek2016}
M.~Řeřábek and T.~Ebrahimi, ``New light field image dataset,'' in \emph{8th International Workshop on Quality of Multimedia Experience (QoMEX)}, Lisbon, Portugal, 2016.

\bibitem{dansereau2019liff}
\BIBentryALTinterwordspacing
D.~G. Dansereau, B.~Girod, and G.~Wetzstein, ``{LiFF}: Light field features in scale and depth,'' in \emph{Computer Vision and Pattern Recognition ({CVPR})}.\hskip 1em plus 0.5em minus 0.4em\relax IEEE, Jun. 2019. [Online]. Available: \url{http://dgd.vision/Papers/dansereau2019liff.pdf}
\BIBentrySTDinterwordspacing

\bibitem{ITURec500}
``Recommendation 500-10: Methodology for the subjective assessment of the quality of television pictures,'' ITU-R Rec. BT.500, 2000.

\bibitem{Li938}
J.~Li, M.~Barkowsky, and P.~Le~Callet, ``Analysis and improvement of a paired comparison method in the application of 3dtv subjective experiment,'' in \emph{2012 19th IEEE International Conference on Image Processing}, 2012, pp. 629--632.

\bibitem{BradleyTerry}
E.~Zermelo, ``Die berechnung der turnier-ergebnisse als ein maximumproblem der wahrscheinlichkeitsrechnung,'' \emph{Mathematische Zeitschrift}, vol.~29, 1929.

\bibitem{LRT}
\BIBentryALTinterwordspacing
\emph{Likelihood Ratio Test}.\hskip 1em plus 0.5em minus 0.4em\relax New York, NY: Springer New York, 2008, pp. 309--316. [Online]. Available: \url{https://doi.org/10.1007/978-0-387-32833-1_233}
\BIBentrySTDinterwordspacing

\bibitem{Wilk_Theorm}
\BIBentryALTinterwordspacing
S.~S. Wilks, ``The large-sample distribution of the likelihood ratio for testing composite hypotheses,'' \emph{The Annals of Mathematical Statistics}, vol.~9, no.~1, pp. 60--62, 1938. [Online]. Available: \url{http://www.jstor.org/stable/2957648}
\BIBentrySTDinterwordspacing

\bibitem{Goodfellow-et-al-2016}
\BIBentryALTinterwordspacing
I.~Goodfellow, Y.~Bengio, and A.~Courville, \emph{Deep Learning}.\hskip 1em plus 0.5em minus 0.4em\relax MIT Press, 2016, book in preparation for MIT Press. [Online]. Available: \url{http://www.deeplearningbook.org}
\BIBentrySTDinterwordspacing

\bibitem{matlab}
\BIBentryALTinterwordspacing
T.~M. Inc., ``Statistics and machine learning toolbox,'' Natick, Massachusetts, United States, 2023. [Online]. Available: \url{https://www.mathworks.com/help/stats/index.html}
\BIBentrySTDinterwordspacing

\end{thebibliography}

\end{document}

\bstctlcite{IEEEexample:BSTcontrol}

\title{
\LARGE
Supplementary Material: DASC: Depth-of-Field Aware Scene Complexity Metric for 3D Visualization on Light Field Display
}

\author{Kamran Akbar\,\orcidlink{0009-0001-8705-8808}, Federica Battisti\,\orcidlink{0000-0002-0846-5879}, Robert Bregovic\,\orcidlink{0000-0002-3878-7588}}

\markboth{Journal of IEEE Transaction on Multimedia}%
{Shell \MakeLowercase{\textit{et al.}}: A Sample Article Using IEEEtran.cls for IEEE Journals}


\maketitle

\section{Dataset}
\label{introduction}
\begin{figure*}[!b]
    \centering
    \includegraphics[width=\textwidth]{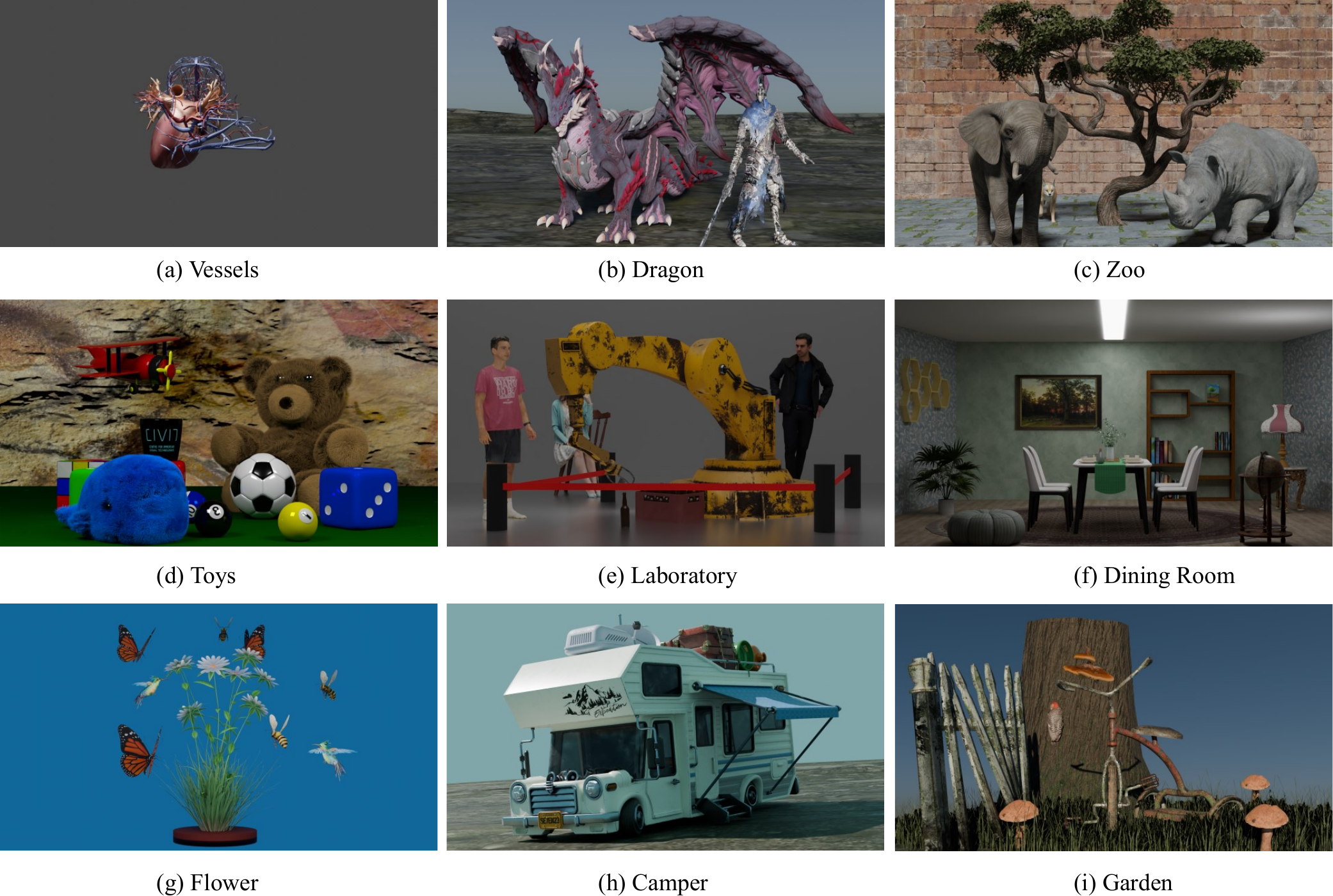}
    \caption{ Rendered middle view of scenes used for subjective studies.}
    \label{fig:scenes}
\end{figure*}

The dataset provided in this study contains nine full-parallax densely sampled light fields. The central view of each light field is shown in Figure \ref{fig:scenes}. The first six scenes (a-f) are used in the \textit{correlation study} and the last three scenes (g-i) are used in the \textit{validation study}. Table I shows the parameters utilized for dataset preparation which is identical for all scenes.

The heterogeneity of the dataset is analyzed with multiple features e.g. \ac{SI}, \ac{CF}, contrast, \ac{TI}, and depth distribution of scene objects. \ac{SI} computes the pixels standard deviation in each view of the light field after applying Sobel filter \cite{characterization}. \ac{CF} and contrast conveys visual information related to perceptual quality and image naturalness \cite{characterization}. In this study, we use \ac{TI} to estimate the difference between two consecutive views in the light field angular domain \cite{ITURec500}. 

\begin{table}[h!]
\caption{Parameters for Light Field Rendering and visualization}
\begin{tabular}{|l|l|l|l|}
\hline
                                    & \textbf{Name}                                                                                       & \textbf{Symbol} & \textbf{Value} \\ \hline
\multirow{9}{*}{\rotatebox[origin=c]{90}{\textbf{Rendering}}} & \begin{tabular}[c]{@{}l@{}}Distance from screen plane to the front\\ part of the scene\end{tabular} & $z_{min}$         & 0.7 m          \\ \cline{2-4} 
                                    & \begin{tabular}[c]{@{}l@{}}Distance from screen plane to the back\\ part of the scene\end{tabular}  & $z_{max}$         & 0.91 m         \\ \cline{2-4} 
                                    & Distance of the viewer plane to the screen                                                          & $z_f$           & 3 m            \\ \cline{2-4} 
                                    & Baseline                                                                                            & $B$             & 2.64 m         \\ \cline{2-4} 
                                    & Distance between adjacent views                                                                     & $b$             & 3.77 mm        \\ \cline{2-4} 
                                    & View resolution                                                                                     & $R_c$           & 1280 px        \\ \cline{2-4} 
                                    & Angular resolution                                                                                  & $\alpha_c$      & 0.072°         \\ \cline{2-4} 
                                    & Circular filter radii range                                                                         & $r$             & 0-15 views     \\ \cline{2-4} 
                                    & Optimal circular filter radius                                                                      & $\hat r$        & 7 views        \\ \hline
\multirow{4}{*}{\rotatebox[origin=c]{90}{\textbf{Display}}}  & Angular resolution                                                                                  & $\alpha_s$      & 0.95°          \\ \cline{2-4} 
                                    & Spatial resolution                                                                                  & $R_s$           & 1280 px        \\ \cline{2-4} 
                                    & DoF range                                                                                           & $d_{\Phi}$      & 0.2 m          \\ \cline{2-4} 
                                    & Field of view                                                                                       & $FoV_s$         & 70°            \\ \hline
\end{tabular}
\end{table}

\begin{figure*}[t!]
    \centering
    \includegraphics[width=\textwidth,keepaspectratio]{images/image1.pdf}
    \caption{ Light field characterization with SI, CF, TI, and Contrast. }
    \label{fig:features_scenes}
\end{figure*}
 \begin{figure*}[t!]
    \centering
    \includegraphics[width=\textwidth]{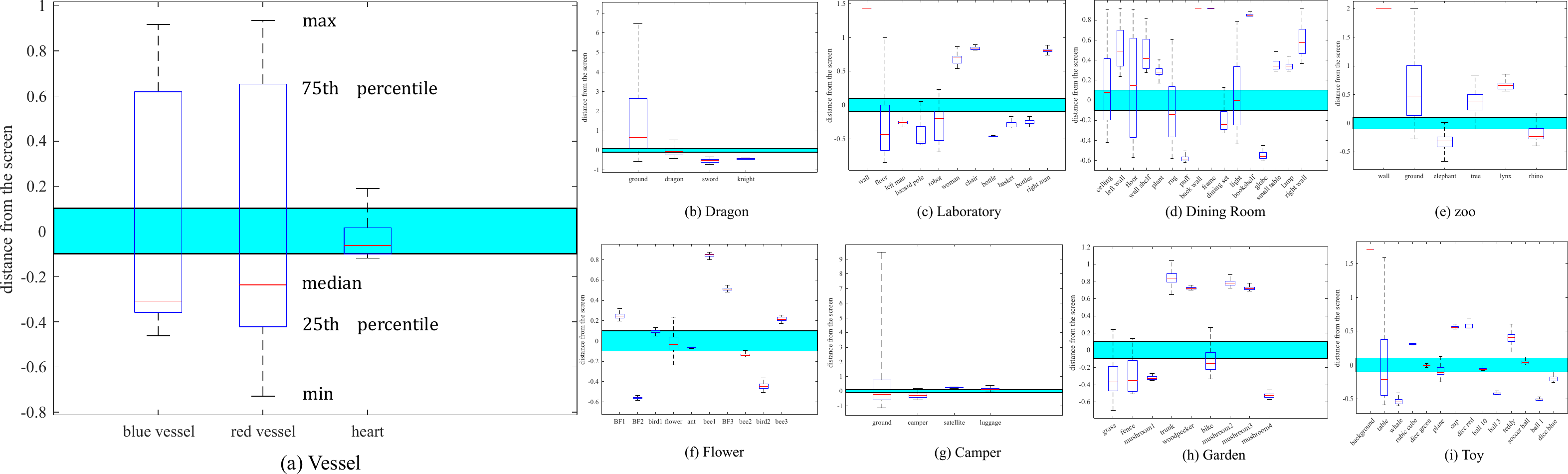}
    \caption{Depth distribution of objects in each scene using a box plot, representing the minimum, maximum, median depth, and the 25th and 75th percentile depths. The shaded region corresponds to the \ac{DoF} of Holovizio 722 RC. }
    \label{fig:ValDepthDist}
\end{figure*}
Figure \ref{fig:features_scenes} shows that  \ac{SI}, \ac{TI}, \ac{CF} and contrast are well-distributed over the span. The depth distribution of each object in the scene provides information about objects placement in the scene with respect to the light field display's \ac{DoF}. Figure \ref{fig:ValDepthDist} illustrates objects' depth distribution for each scene in the dataset. The shaded region is the \ac{DoF} of Holovizio 722 RC.

\bibliographystyle{IEEEtran}
\bibliography{bibliography}